\documentclass[aps,prd,preprint,showpacs,nofootinbib,superscriptaddress]{revtex4-1}

\usepackage[utf8]{inputenc}
\usepackage[T1]{fontenc}

\usepackage{amsmath,amssymb,amsfonts}
\usepackage{braket}

\usepackage{graphicx}
\usepackage{tikz}
\usetikzlibrary{arrows.meta,positioning,fit,calc}

\usetikzlibrary{spy,calc}

\begin{document}

\title{Topological Relics of Infrared QED in the Rayleigh–Jeans Regime}

\author{J. Gamboa}
\email{jorge.gamboa@usach.cl}
\affiliation{Departamento de F\'isica, Universidad de Santiago de Chile, Santiago 9170020, Chile}


\begin{abstract}
We revisit infrared Quantum Electrodynamics (QED) by reformulating the dressing of charged states in terms of a functional Berry connection. In this framework, the electron--photon cloud acquires a topological character, leading to a modification of the Rayleigh--Jeans limit of blackbody radiation. The deviation is controlled by a dimensionless parameter $\epsilon$, estimated as $\epsilon \simeq (1.7$--$2.3)\times 10^{-3}$ in the $80$--$110~\mathrm{GHz}$ range, corresponding to a fractional shift of $\sim 0.2\%$. This effect lies within the sensitivity of current high--precision radiometry and, being system--independent, should be observable both in laboratory blackbody spectra and in the cosmic microwave background.
\end{abstract}

\maketitle

Quantum Electrodynamics (QED) in the infrared regime is a subtle subject, not only because it requires a well-defined prescription for the cancellation of infrared divergences, but also because one must understand in detail why such divergences cancel in the first place. Since the pioneering work of Bloch and Nordsieck~\cite{BN}, the problem has been revisited by many authors, who have refined and formalized the original idea~\cite{varios}. Among the most influential contributions are those of Chung~\cite{chung}, Kibble~\cite{kibble}, and Kulish and Faddeev~\cite{KF}, who introduced the concept of dressed states. While this approach ensures the cancellation of infrared divergences, the exact structure of the dressing, its physical origin, and the interpretation of the resulting electron--photon clouds have remained to a large extent unclear.

A new perspective emerged with the introduction of the Berry phase in the early 1980s \cite{berry,shapere,review} whose deep connection with the adiabatic and Born-Oppenheimer approximations opened the door to reformulating the dressing problem in geometric and topological terms. Within this framework, the dressing can be understood as a phase of topological origin, arising naturally in the adiabatic evolution of the system \cite{gamboa}.
In the adiabatic approximation (and within the Born-Oppenheimer framework), the key outcome is the emergence of a purely geometric gauge symmetry -not associated with dynamical degrees of freedom- whose presence gives rise to geometric and topological properties that become explicit only in the low-energy limit.

In the infrared regime of QED, physical states no longer belong to the Fock space. Constructing states in this non-perturbative sector has been the focus of mathematically rigorous quantum field theory \cite{strocchi, Buchholz1} . In our approach, we employ formal \emph{dressed states} in the sense discussed in  \cite{strocchi, Buchholz1} defined as  
\[
\ket{\mathrm{in}} \;\longrightarrow\; \hat{U}(C)\,\ket{\mathrm{in}}, 
\qquad 
\ket{\mathrm{out}} \;\longrightarrow\; \hat{U}(C)\,\ket{\mathrm{out}},
\]
where the dressing operator is  
\[
\hat{U}(C) \;=\; \exp\!\left[i\,\gamma_5 \oint_C dx^\mu\, \mathcal{A}_\mu\right].
\]
Here, \(C\) denotes a path in the space of gauge configurations, and \(\mathcal{A}_\mu\) is a \(U(1)\) Berry-type functional connection generated by adiabatic transport in the IR sector. In our construction, we fix the \emph{adiabatic gauge} by imposing  
\[
\mathcal{A}_\mu = i\,\gamma_5\,\partial_\mu \alpha,
\]
so that \(\mathcal{A}_\mu\) directly encodes the coherent photon cloud dressing the charged particles, ensuring gauge invariance of the asymptotic states and capturing the topological and geometric properties that survive in the long-wavelength limit.  

In the chiral basis, the dressing operator \({\hat U}(C)\) acts as  
\[
{\hat U}(C) = \cos\!\Phi[C]\,\mathbb{I} + i\,\sin\!\Phi[C]\,\gamma_5
=
\begin{pmatrix}
e^{i\Phi[C]} & 0 \\
0 & e^{-i\Phi[C]}
\end{pmatrix},
\qquad
\Phi[C] := \oint_C dx^\mu\,\mathcal{A}_\mu.
\]
If the dressed state is required to be fermionic, the Berry flux must be quantized as  
\[
\Phi[C] = (2n+1)\,\pi, \qquad n \in \mathbb{Z},
\]
in which case \({\hat U}(C) = -\mathbb{I}\), producing the global fermionic sign.

This quantization of the functional flux has significant physical implications, as the cloud is a bound state that effectively acts as the \emph{quantum} replacing the conventional photon in the infrared sector of blackbody radiation. In other words, the cloud can be viewed as the ontological manifestation of an elementary quantum in the infrared regime.

\begin{figure}[ht]
    \centering
    \IfFileExists{modified_RJ_T300_70to110GHz_highres.png}{%
      \includegraphics[width=0.85\textwidth]{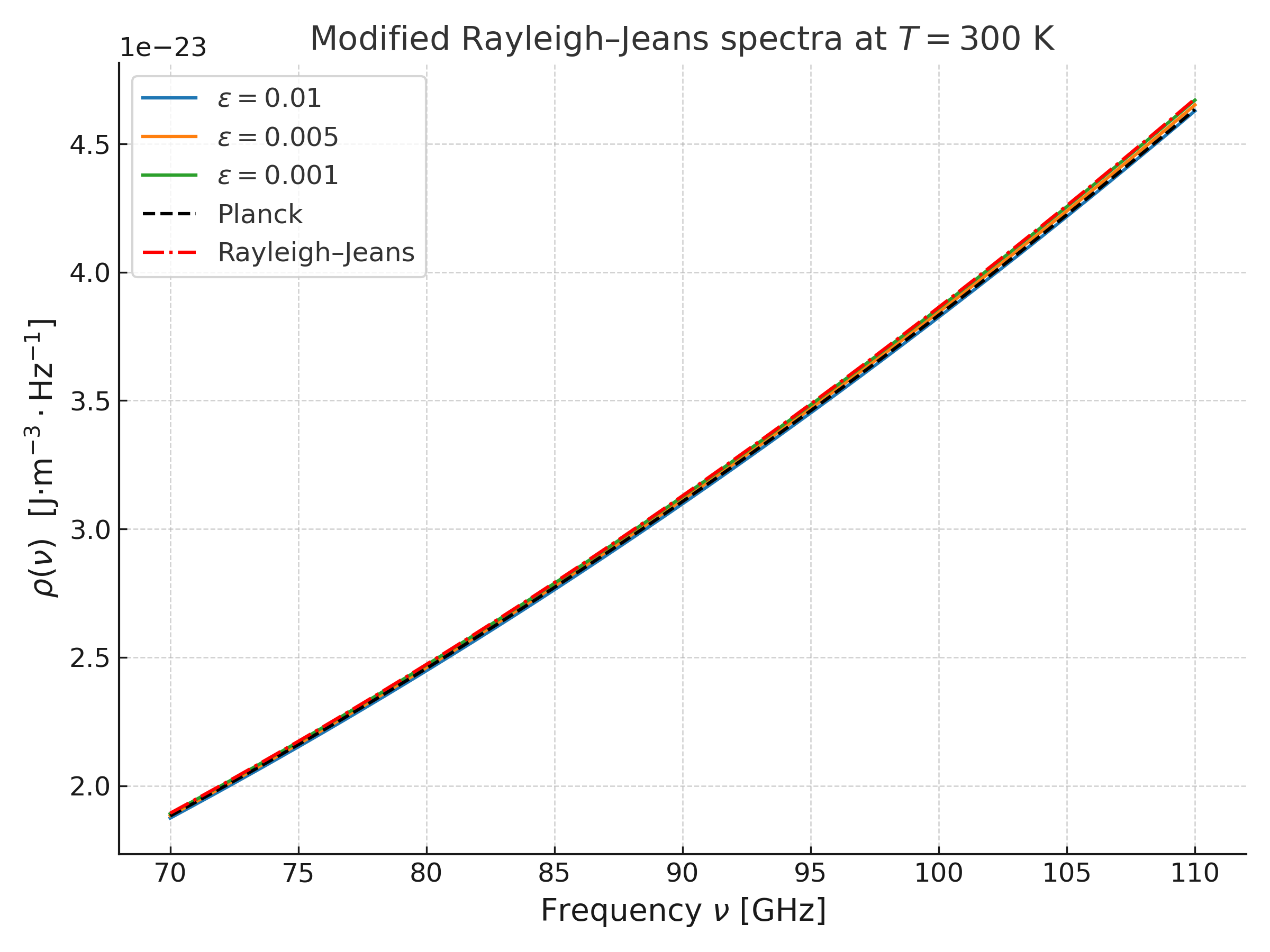}%
    }{%
      \fbox{%
        \begin{minipage}[c][45mm][c]{0.85\textwidth}
        \centering
        [Placeholder: falta \texttt{modified\_RJ\_T300\_70to110GHz\_highres.png}]
        \end{minipage}%
      }%
    }
    \caption{Modified spectra at $T = 300~\mathrm{K}$ in the range $10$--$200~\mathrm{GHz}$ 
    for different values of $\epsilon$ (flux quantized). 
    The Planck spectrum, the Rayleigh--Jeans limit, and the modified curves are shown for comparison.}
    \label{fig:modified_RJ_T300_no_lambda}
\end{figure}

Moreover, since the cloud constitutes the relevant quantum for electromagnetic phenomena in the infrared, its presence should modify the Rayleigh--Jeans law. Such deviations are, in principle, observable both in controlled laboratory experiments and in the Cosmic Microwave Background (CMB)~\cite{CMB1,CMB2}.

This modification follows directly from the corrections to Planck’s distribution law derived in Appendix~A. In the infrared limit, the Rayleigh–Jeans law is accordingly modified to  
\[
\rho_{\rm RJ}(\nu) \;=\; \frac{8\pi\nu^{2}}{c^{3}}\,k_{B}T\,\big(1 - \epsilon\big),
\]
where the dimensionless parameter \(\epsilon\) quantifies the relative amplitude of the cloud effect. By comparing this expression with the standard Rayleigh–Jeans law, one finds  
\[
\frac{\rho - \rho_{\rm RJ}}{\rho_{\rm RJ}} \;=\; \epsilon  ,
\]
For small values of $\epsilon$, this expression shows that $\epsilon$ directly measures the relative deviation from the classical Rayleigh-Jeans spectrum.

Figure~\ref{fig:modified_RJ_T300_no_lambda} shows the deviations from the Rayleigh--Jeans law for \(\epsilon\) in the range \(10^{-2}\)–\(10^{-3}\) at \(T=300~\mathrm{K}\) over the frequency interval \(80\)–\(110~\mathrm{GHz}\). The intrinsic difference between the Rayleigh-Jeans and Planck predictions is present across the entire band, increasing smoothly with frequency from about \(0.64\%\) 
at \(80~\mathrm{GHz}\) to \(0.88\%\) at \(110~\mathrm{GHz}\). Detecting an additional signal due to \(\epsilon\) -which acts as an almost constant multiplicative factor on the distribution- requires noise and systematic errors to be reduced below \(\epsilon/3\). In practice, the \(80\)–\(100~\mathrm{GHz}\) region offers a favorable balance between received power and instrumental control, while the \(100\)–\(110~\mathrm{GHz}\) range benefits from a slightly larger Rayleigh--Jeans–Planck signal, facilitating calibration checks.

In summary, our conclusions are as follows. (i) In the optimal frequency range of $80$--$90~\mathrm{GHz}$, the radiometric precision achievable with a well-calibrated hot-cold system ($\sim 10^{-4}$ in relative brightness) is about one hundred times smaller than the expected signal for $\epsilon = 10^{-2}$ and roughly an order of magnitude smaller for $\epsilon = 10^{-3}$. This ensures a robust signal-to-noise ratio in both cases. Given the estimated value $\epsilon \simeq 2\times 10^{-3}$ inferred from the binding-energy argument above, the predicted deviation lies well within the detectable range of current high-precision radiometric techniques \cite{CMB1,CMB2,CMB3}. (ii) In the infrared regime ($h\nu \ll k_{B}T$), the correction to the Rayleigh-Jeans law can be parametrized by a dimensionless constant $\epsilon$ which, to leading order, is independent of both frequency and temperature. This universality implies that the same small deviation should occur in any blackbody spectrum, regardless of its temperature. Consequently, such infrared corrections are expected to be observable not only in the cosmic microwave background, but also in laboratory blackbody experiments at room temperature \cite{CMB3}.

This research was supported by DICYT (USACH), grant number 042531GR$\_$REG.

\appendix
\section{Modified Planck Distribution}
\label{app:modified-planck}
We show how a functional Berry phase accumulated by electromagnetic radiation modes can induce an \emph{oscillatory} correction to Planck's law for the blackbody spectrum. Starting from the expectation value of the energy operator, we analyze the coherent contribution between states \(\ket{n}\) and \(\ket{m}\), modulated by a functional phase \(\Delta \alpha_n\). Under certain approximations, we obtain a corrected spectral energy density of the form:
\begin{equation}
\rho(\nu) = \rho_{\text{Planck}}(\nu) \left[1 + \epsilon \cos\left( \Delta \alpha(\nu) \right)\right],
\end{equation}
where \(\epsilon \ll 1\) and \(\Delta \alpha(\nu)\) is a frequency-dependent phase.

We consider a quantized electromagnetic field inside a cavity. The energy operator for a mode of frequency \(\nu\) is given by:
\begin{equation}
\hat{E}_\nu = h\nu\, a^\dagger_\nu a_\nu.
\end{equation}

Suppose the system is in a state expressed as a superposition in the number basis:
\begin{equation}
\ket{\Psi} = \sum_n c_n\, e^{i \Delta \alpha_n} \ket{n},
\end{equation}
where \(\Delta \alpha_n\) is a functional phase that depends on the mode and is related to infrared coupling and the functional Berry connection. Importantly, in the adiabatic approximation we can use:
\begin{equation}
\Delta \alpha_n \approx n\, \Delta \alpha.
\end{equation}

The expectation value of the energy is then:
\begin{align}
\langle \hat{E}_\nu \rangle 
&= \sum_{n,m} c_n^* c_m\, e^{-i\Delta \alpha_n} e^{i\Delta \alpha_m} 
\bra{n} h\nu\, a^\dagger_\nu a_\nu \ket{m} \notag \\
&= h\nu \sum_n |c_n|^2\, n 
+ h\nu \sum_{\substack{n,m \\ n\neq m}} c_n^* c_m\,
\bra{n} a^\dagger_\nu a_\nu \ket{m}\,
e^{i(\Delta \alpha_m - \Delta \alpha_n)}.
\end{align}

The second sum contains coherent terms, which typically vanish in thermal distributions. 
However, if partial coherence exists between \(\ket{n}\) and \(\ket{n+1}\), then
\begin{equation}
\bra{n} a^\dagger_\nu a_\nu \ket{n+1} 
= \sqrt{n+1}\,\delta_{m,n+1}.
\end{equation}

This leads to a contribution of the form:
\begin{align}
\langle \hat{E}_\nu \rangle 
&\approx h\nu \sum_n |c_n|^2\, n 
+ h\nu \sum_n \Re\!\left( c_n^* c_{n+1}\, \sqrt{n+1}\, e^{i(\Delta \alpha_{n+1} - \Delta \alpha_n)} \right).
\end{align}

Assuming the functional phase varies smoothly with \(n\), we can write:
\begin{equation}
\Delta \alpha_{n+1} - \Delta \alpha_n \approx \Delta \alpha(\nu),
\end{equation}
and suppose \( c_n^* c_{n+1} \sqrt{n+1} \) is real and positive (quasi-coherent state). Then:
\begin{equation}
\langle \hat{E}_\nu \rangle \approx \langle \hat{E}_\nu \rangle_{\text{Planck}} \left[1 + \epsilon \cos\left( \Delta \alpha(\nu) \right)\right],
\end{equation}
where \(\epsilon \ll 1\) quantifies the relative amplitude of the oscillations.

The spectral energy density is defined as:
\begin{equation}
\rho(\nu) = \frac{\langle \hat{E}_\nu \rangle}{V\, d\nu},
\end{equation}
and for an ideal blackbody it is given by Planck’s law:
\begin{equation}
\rho_{\text{Planck}}(\nu) = \frac{8\pi h \nu^3}{c^3} \frac{1}{e^{h\nu/k_B T} - 1}.
\end{equation}

Therefore, the functional correction can be written as:
\begin{equation}
\rho(\nu) = \rho_{\text{Planck}}(\nu) \left[1 + \epsilon \cos\left( \Delta \alpha(\nu) \right)\right],
\end{equation}
where \(\Delta \alpha(\nu)\) typically takes the form:
\begin{equation}
\Delta \alpha(\nu) = \frac{2\pi \nu}{\Lambda_{\text{IR}}} + \phi,
\end{equation}
with \(\Lambda_{\text{IR}}\) an infrared scale associated with the functional coupling, and \(\phi\) a constant phase.

The presence of functional phases associated with the Berry connection in radiation modes can induce oscillations on top of Planck’s law. These oscillations constitute a possible observable signature of nonperturbative infrared effects in gauge theories.

\end{document}